\documentclass[twocolumn,superscriptaddress,showpacs,preprintnumbers,amsmath,amssymb]{revtex4}

\usepackage{graphicx,color}
\usepackage{dcolumn}
\usepackage{bm}
\usepackage{mathrsfs}
\usepackage{amsmath}

\DeclareMathOperator{\arcsinh}{arcsinh}

\newcommand{\lb}{\left(}
\newcommand{\rb}{\right)}
\newcommand{\ls}{\left[}
\newcommand{\rs}{\right]}

\begin{document}

\title{The Slater approximation for Coulomb exchange effects in nuclear covariant density functional theory}

\author{Huai-Qiang Gu}
 \affiliation{School of Nuclear Science and Technology, Lanzhou University,
    Lanzhou 730000, China}
 \affiliation{School of Physics and Nuclear Energy Engineering, Beihang University,
    Beijing 100191, China}

\author{Haozhao Liang\footnote{Email: haozhao.liang@riken.jp}}
 \affiliation{RIKEN Nishina Center, Wako 351-0198, Japan}
 \affiliation{State Key Laboratory of Nuclear Physics and Technology, School of Physics,
    Peking University, Beijing 100871, China}

\author{Wen Hui Long}
 \affiliation{School of Nuclear Science and Technology, Lanzhou University,
    Lanzhou 730000, China}

\author{Nguyen Van Giai}
 \affiliation{Institut de Physique Nucl\'eaire, IN2P3-CNRS and Universit\'e Paris-Sud,
    F-91406 Orsay Cedex, France}

\author{Jie Meng}
 \affiliation{State Key Laboratory of Nuclear Physics and Technology, School of Physics,
    Peking University, Beijing 100871, China}
 \affiliation{School of Physics and Nuclear Energy Engineering, Beihang University,
    Beijing 100191, China}
 \affiliation{Department of Physics, University of Stellenbosch, Stellenbosch, South Africa}

\date{\today}

\begin{abstract}
The relativistic local density approximation (LDA) for the Coulomb
exchange functional in nuclear systems is presented. This
approximation is composed of the well-known Slater approximation in
the non-relativistic scheme and the corrections due to the
relativistic effects. Its validity in finite nuclei is examined by
comparing with the exact treatment of the Coulomb exchange term in
the relativistic Hartree-Fock-Bogoliubov theory. The relativistic
effects are found to be important and the exact Coulomb exchange
energies can be reproduced by the relativistic LDA within $5\%$
demonstrated by the semi-magic Ca, Ni, Zr, Sn, and Pb isotopes from
proton drip line to neutron drip line.
\end{abstract}

\pacs{
21.60.Jz, 
24.10.Jv, 
21.10.Sf, 
31.15.eg  
}
\maketitle


The Coulomb interaction between protons is one of the most important building blocks in atomic nuclei \cite{Nolen1969,Auerbach1983}.
It is the main source of the charge symmetry-breaking in the nuclear Hamiltonian.
Globally, the Coulomb interaction is the driving force for determining the $\beta$-stability valley away from the $N=Z$ line.
Specifically, the Coulomb interaction plays a crucial role in understanding the Coulomb displacement energies, isospin mixing, proton emission, fission barriers, $\alpha$-decay energy release $Q_\alpha$ in superheavy elements, and more.
Therefore, the nuclear density functional theory (DFT) that models the effective strong interactions has to be accompanied by an explicit functional form for the Coulomb interaction \cite{Bender2003}.

Generally speaking, the Coulomb interaction is the best known part of the nuclear Hamiltonian, and both its direct (Hartree) and exchange (Fock) terms can be exactly calculated at the mean-field level.
However, the exchange term is very cumbersome to include due to the non-locality of the corresponding mean field, especially for the deformed nuclei.
In the spirit of the DFT of Kohn and Sham \cite{Kohn1965}, the local density approximation (LDA) has been intensively adopted in the calculations of finite nuclei since the 1970s \cite{Negele1970}.
So far, in almost all of the non-relativistic (NR) self-consistent Hartree-Fock calculations, the Coulomb exchange energy and the corresponding single-particle potential are evaluated within a local scheme by using the so-called Slater approximation \cite{Slater1951}.
The validity of this approximation has been discussed in the Skyrme \cite{Titin-Schnaider1974,Skalski2001,LeBloas2011} and Gogny \cite{Anguiano2001} approaches.

During the past decades, the nuclear covariant density functional
theory (CDFT) at the Hartree level, also known as the relativistic
Hartree (RH) or relativistic mean-field (RMF) theory, has received
much attention due to its successful description of many nuclear
phenomena
\cite{Ring1996,Lalazissis2004,Vretenar2005,Meng2006,Paar2007,Niksic2011,Meng2011}.
The Lorentz covariant form of the theory itself guarantees the
self-consistent spin-orbit potential and also puts stringent
restrictions on the number of parameters in the corresponding
functionals without reducing the quality of the agreement with
experimental data. In order to retain the required simplicity in
the RH framework, the non-local Coulomb exchange term is usually
neglected.
Its effects on the binding energies, radii, etc., are assumed to be absorbed into the effective meson-nucleon coupling strengths through the phenomenological fit of the model.
However, this prescription of neglecting the Coulomb exchange term is not always valid, because the meson-nucleon interactions are of isospin symmetry but the Coulomb interaction is not.
One
example is the isospin symmetry-breaking corrections to the
superallowed $\beta$ decays \cite{Liang2009}, which are crucial for
testing the unitarity of the Cabibbo-Kobayashi-Maskawa matrix
\cite{Towner2010}.

Recently, the density-dependent relativistic Hartree-Fock (RHF) theory \cite{Long2006a,Long2007} has been developed and achieved equivalent success in describing the ground- and excited-state properties as the RH theory.
It has been shown that the meson exchange terms play very important roles in the nucleon effective mass splitting \cite{Long2006a}, symmetry energies \cite{Sun2008}, spin and pseudospin symmetries \cite{Long2006b,Liang2010}, shell structure and its evolutions \cite{Long2007,Long2008,Tarpanov2008,Long2009,Moreno-Torres2010}, deformation \cite{Ebran2011}, and spin-isospin resonances \cite{Liang2008,Liang2012a}.
Subsequently, a unified and self-consistent description of both RHF mean field and pairing correlations, i.e., the relativistic Hartree-Fock-Bogoliubov (RHFB) theory \cite{Long2010a,Long2010b}, has been achieved for the exotic nuclei far from the $\beta$-stability valley.
In the frameworks of RHF and RHFB, for the first time, the non-local Coulomb exchange term has been taken into account exactly, which is crucial for understanding the isospin corrections to the superallowed $\beta$ decays \cite{Liang2009}.

With the simplicity of the local CDFT and the success in treating the Coulomb terms exactly, it is worthwhile to explore the relativistic LDA for the Coulomb exchange term, a topic less known in the nuclear physics community.
The relativistic LDA may be a simple approach to implement the missing Coulomb exchange effects in the RH theory, yet keeping the merits of locality.
The attempt in this direction is further supported by a recent success in folding the meson exchange terms into a local equivalent scheme \cite{Liang2012b}.

The relativistic LDA has been explored for electronic systems in past decades. The extension of the Hohenberg and Kohn theorem
\cite{Hohenberg1964} to relativistic systems was first formulated in
Ref.~\cite{Rajagopal1973} by utilizing a quantum electrodynamics
(QED)-based Hamiltonian with the four-current. Within the no-sea
approximation, which is also called no-pair approximation in atomic
physics for neglecting all effects due to the creation of
particle-antiparticle pairs, the total energy of the system can be
expressed as a functional with respect to the four-current
$j^\mu(\mathbf{r})=(\rho(\mathbf{r}), \mathbf{j}(\mathbf{r}))$
\cite{Rajagopal1978,MacDonald1979}, instead of only
$\rho(\mathbf{r})$ in the NR case. In complete analogy with the NR
case, the relativistic LDA for the Coulomb exchange energy
functional $E_{\rm Cex}[j^\mu(\mathbf{r})]$ is constructed by using
the model of relativistic homogeneous electron gas. As the
space-like component $\mathbf{j}$ vanishes in the homogeneous
systems, the full $j^\mu$-dependence of the exact exchange energy
functional is reduced to a pure density-dependence in the
relativistic LDA. The relativistic corrections, i.e., the
differences between $E^{\rm RLDA}_{\rm Cex}[\rho(\mathbf{r})]$ and
$E^{\rm NRLDA}_{\rm Cex}[\rho(\mathbf{r})]$, are shown to be
substantial, see Ref.~\cite{Engel2011} for details.

In this Rapid Communication, the relativistic LDA for the Coulomb interaction in nuclear physics will be presented.
The main focus will be its validity by comparing with the exact RHFB results and the relativistic corrections to the traditional Slater approximation.

%

Before introducing the relativistic LDA, it is illuminating to briefly recall the Coulomb energies and the corresponding LDA for the exchange term in the NR systems.

In the non-relativistic scheme, the proton density distribution in nuclei reads
\begin{equation}
    \rho_p(\mathbf{r}) = \sum_i^p v_i^2 \psi^*_i(\mathbf{r})\psi_i(\mathbf{r}),
\end{equation}
with the single-particle wave functions $\psi_i(\mathbf{r})$ and occupation probabilities $v_i^2$.
The direct term of Coulomb energy is simply a functional of $\rho_p(\mathbf{r})$, i.e.,
\begin{equation}\label{Eq:EdirNR}
    E_{\rm Cdir}=\frac{e^2}{2}\int\int
    d^3r d^3r'\frac{\rho_p(\mathbf{r})\rho_p(\mathbf{r}')}{|\mathbf{r}-\mathbf{r}'|},
\end{equation}
and the corresponding Hartree potential for protons reads
\begin{equation}
    V_{\rm Cdir}(\mathbf{r})=\frac{\delta E_{\rm Cdir}[\rho_p(\mathbf{r})]}{\delta \rho_p(\mathbf{r})}
    =e^2\int
    d^3r'\frac{\rho_p(\mathbf{r}')}{|\mathbf{r}-\mathbf{r}'|}.
\end{equation}

In contrast, the exchange term of Coulomb energy,
\begin{equation}\label{Eq:EexNR}
    E_{\rm Cex} = -\frac{e^2}{2}\sum_{ij}^pv_i^2v_j^2\int\int d^3r d^3r'
        \frac{\psi^*_i(\mathbf{r})\psi_j(\mathbf{r})\psi^*_j(\mathbf{r}')\psi_i(\mathbf{r}')}{|\mathbf{r}-\mathbf{r}'|},
\end{equation}
leads to non-local single-particle potentials and thus is more complicated to calculate.

In the spirit of the LDA, the exchange energy density of an inhomogeneous system with density $\rho(\mathbf{r})$ is locally approximated by the exchange energy density of a homogeneous system with density $n = \rho(\mathbf{r})$.
In the homogeneous nuclear matter, the single-particle wave functions are the plane-wave solutions of the Schr\"odinger equation, so that the Coulomb exchange energy shown in Eq.~(\ref{Eq:EexNR}) per unit volume can be calculated analytically as
\begin{equation}\label{Eq:exNR}
    e_{\rm Cex} = -\frac{3}{4}\lb\frac{3}{\pi}\rb^{1/3}e^2n_p^{4/3},
\end{equation}
where $n_p$ is the proton density.
Therefore, by using the LDA, the Coulomb exchange energy in finite nuclei is determined as
\begin{equation}\label{Eq:ELDA}
    E^{\rm LDA}_{\rm Cex}
    = -\frac{3}{4}\lb\frac{3}{\pi}\rb^{1/3}e^2\int d^3r\rho_p^{4/3}(\mathbf{r}).
\end{equation}
This is the well-known Slater approximation \cite{Slater1951} widely used in the non-relativistic DFT.
The corresponding single-particle potential for protons reads
\begin{equation}\label{Eq:VLDA}
    V^{\rm LDA}_{\rm Cex}(\mathbf{r})=\frac{\delta E^{\rm LDA}_{\rm Cex}[\rho_p(\mathbf{r})]}{\delta \rho_p(\mathbf{r})}
    =-\left(\frac{3}{\pi}\right)^{1/3}e^2\rho_p^{1/3}(\mathbf{r}).
\end{equation}

In the relativistic scheme, the variables of the energy functional include not only the density distributions but also the currents.
The proton density distribution and currents read
\begin{subequations}
\begin{eqnarray}
    \rho_p(\mathbf{r}) &=& \sum_i^p v_i^2 \bar\psi_i(\mathbf{r})\gamma^0\psi_i(\mathbf{r}),\\
    \mathbf{j}_p(\mathbf{r}) &=& \sum_i^p v_i^2 \bar\psi_i(\mathbf{r})\boldsymbol{\gamma}\psi_i(\mathbf{r}),
\end{eqnarray}
\end{subequations}
respectively.
It should be emphasized that the summations over $i$ include only the particles in the Fermi sea, i.e., the no-sea approximation, which is the so-called no-pair approximation in atomic physics \cite{Engel2011}.

Correspondingly, the direct term of Coulomb energy reads
\begin{eqnarray}\label{Eq:EdirR}
    E^{\rm R}_{\rm Cdir}&=&\frac{e^2}{2}\int\int
    d^3r d^3r'\frac{j_p^\mu(\mathbf{r})j_{p,\mu}(\mathbf{r}')}{|\mathbf{r}-\mathbf{r}'|}\nonumber\\
    &=& \frac{e^2}{2}\int\int
    d^3r d^3r'\ls\frac{\rho_p(\mathbf{r})\rho_p(\mathbf{r}')}{|\mathbf{r}-\mathbf{r}'|}
    - \frac{\mathbf{j}_p(\mathbf{r})\cdot \mathbf{j}_p(\mathbf{r}')}{|\mathbf{r}-\mathbf{r}'|}\rs,\nonumber\\
\end{eqnarray}
where the first and second terms are the contributions from the time-like and space-like components of the Coulomb field, respectively.
They are called Coulomb and transverse contributions in Ref.~\cite{Engel2011}.
For the systems with time-reversal symmetry, the contribution from the space-like component vanishes.
Then, Eq.~(\ref{Eq:EdirR}) has the same structure as Eq.~(\ref{Eq:EdirNR}).

On the other hand, the exchange term of Coulomb energy is \cite{Engel2011}
\begin{eqnarray}\label{Eq:EexR}
    E^{\rm R}_{\rm Cex}&=&-\frac{e^2}{2}\sum_{ij}^pv_i^2v_j^2\int\int d^3r d^3r'
    \frac{\cos(|\varepsilon_i-\varepsilon_j||\mathbf{r}-\mathbf{r}'|)}{|\mathbf{r}-\mathbf{r}'|}\nonumber\\
    &&\qquad\times\bar\psi_i(\mathbf{r})\gamma^\mu\psi_j(\mathbf{r})\bar\psi_j(\mathbf{r}')\gamma_\mu\psi_i(\mathbf{r}'),
\end{eqnarray}
where $\varepsilon_i$ are the single-particle energies.

In the relativistic homogeneous nuclear matter, the single-particle wave functions are the plane-wave solutions of the Dirac equation.
As a result, the time-like component $\bar{e}^{\rm R}_{\rm Cex}$ and space-like component $\bar{\bar{e}}^{\rm R}_{\rm Cex}$ of the Coulomb exchange energy shown in Eq.~(\ref{Eq:EexR}) per unit volume can be, respectively, expressed as  \cite{MacDonald1979,Engel2011}
\begin{equation}\label{Eq:exR}
    \bar{e}^{\rm R}_{\rm Cex}= e_{\rm Cex} \bar\Phi(\beta)\quad\mbox{and}\quad
    \bar{\bar{e}}^{\rm R}_{\rm Cex}= e_{\rm Cex} \bar{\bar\Phi}(\beta),
\end{equation}
by the NR one $e_{\rm Cex}$ in Eq.~(\ref{Eq:exNR}) together with
\begin{subequations}\label{Eq:Phi}
\begin{eqnarray}
    \bar\Phi(\beta)&=&\frac{5}{6}+\frac{1}{3\beta^2}+\frac{2\eta}{3\beta}\arcsinh\beta
        -\frac{2\eta^4}{3\beta^4}\ln\eta\nonumber\\
    &&-\frac{1}{2}\lb\frac{\eta}{\beta}-
        \frac{\arcsinh\beta}{\beta^2}\rb^2 \nonumber\\
    &=&1-\frac{1}{9}\beta^2+\frac{13}{180}\beta^4+\cdots\\
    \bar{\bar\Phi}(\beta)&=&\frac{1}{6}-\frac{1}{3\beta^2}-\frac{2\eta}{3\beta}\arcsinh\beta
        +\frac{2\eta^4}{3\beta^4}\ln\eta\nonumber\\
    &&-\lb\frac{\eta}{\beta}-
        \frac{\arcsinh\beta}{\beta^2}\rb^2 \nonumber\\
    &=&-\frac{5}{9}\beta^2+\frac{59}{180}\beta^4+\cdots,
\end{eqnarray}
\end{subequations}
where $\beta=(3\pi^2n_p)^{1/3}/M$, $\eta=\sqrt{1+\beta^2}$, and $M$ is the proton mass.
The details can be found in Ref.~\cite{Engel2011} and the references therein.

From Eqs.~(\ref{Eq:exR}) and (\ref{Eq:Phi}), the relativistic
results are identical to the NR counterparts at the zero density
limit.
The relativistic corrections to the Coulomb exchange energy increase with the density.
For the symmetric nuclear matter, $n_p\approx0.08$~fm$^{-3}$ at
the saturation density, one has $\beta\sim0.28$,
$\beta^2\sim0.078$, and $\beta^4\sim0.006$. Therefore, the
relativistic corrections to the Coulomb exchange energy are expected
to be substantial in nuclear systems, and the contributions from
$\beta^4$ and higher-order terms can be neglected. Accordingly, the
relativistic effect from the space-like component is 5 times that
from the time-like component, as shown in Eq. (\ref{Eq:Phi}).

From Eq.~(\ref{Eq:exR}), up to the order of $\beta^2$, the Coulomb
exchange energy in the relativistic LDA is
\begin{equation}\label{Eq:ERLDA}
    E^{\rm RLDA}_{\rm Cex}
    = -\frac{3}{4}\lb\frac{3}{\pi}\rb^{1/3}e^2\int d^3r\rho_p^{4/3}
        \ls 1-\frac{2}{3}\frac{(3\pi^2\rho_p)^{2/3}}{M^2}\rs.
\end{equation}
The corresponding contribution to the single-particle potential for protons reads
\begin{equation}\label{Eq:VRLDA}
    V^{\rm RLDA}_{\rm Cex}(\mathbf{r})=
    -\left(\frac{3}{\pi}\right)^{1/3}e^2\rho_p^{1/3}(\mathbf{r})+\lb\frac{3\pi}{M^2}\rb e^2\rho_p(\mathbf{r}).
\end{equation}

%

In the following, the properties of finite nuclei are calculated in
the density-dependent RHFB theory \cite{Long2010a} with PKA1
\cite{Long2007} and D1S \cite{Berger1991} effective interactions for the
particle-hole and particle-particle channels, respectively. The RHFB equations are
solved on a Dirac Woods-Saxon basis \cite{Zhou2003} within a
spherical box of radius $R_{\rm max} = 20$~fm, and the numbers of
positive and negative energy levels for each $(l,j)$-state are fixed
to $N_F = 28$ and $N_D=12$, respectively.

In the standard RHFB calculations, the Coulomb exchange energy and
the corresponding non-local single-particle potential are calculated
exactly. They are labeled with the superscript ``exact". The
corresponding results obtained by using the relativistic LDA shown
in Eqs.~(\ref{Eq:ERLDA}) and (\ref{Eq:VRLDA}) are labeled with the
superscript RLDA. In order to quantitatively investigate the
relativistic corrections in Eq.~(\ref{Eq:exR}), the results of the
non-relativistic LDA shown in Eqs.~(\ref{Eq:ELDA}) and (\ref{Eq:VLDA}) labeled
with the superscript NRLDA are also presented for comparison. The
relative deviations of the approximate Coulomb exchange energies
from the RHFB results are defined as
\begin{equation}\label{Eq:DE}
  \Delta E_{\rm Cex} = \frac{E^{\rm LDA}_{\rm Cex}-E^{\rm exact}_{\rm Cex}}{E^{\rm exact}_{\rm Cex}}.
\end{equation}

\begin{figure}
\centering
\includegraphics[width=8cm]{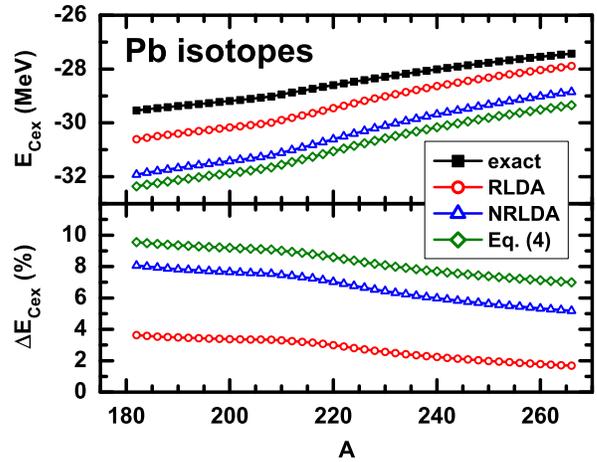}
\caption{(Color online) Coulomb exchange energies in Pb isotopes calculated by RHFB theory \cite{Long2010a} with PKA1 \cite{Long2007}. The results obtained with the same proton density distributions $\rho_p(r)$ but within the relativistic and non-relativistic (NR) local density approximations (LDA) and their relative deviations from the exact results are shown in the upper and lower panels, respectively.
The corresponding results calculated by Eq.~(\ref{Eq:EexNR}), the exact formula for the NR case, are also shown for comparison.
    \label{Fig1}}
\end{figure}

\begin{figure}
\centering
\includegraphics[width=8cm]{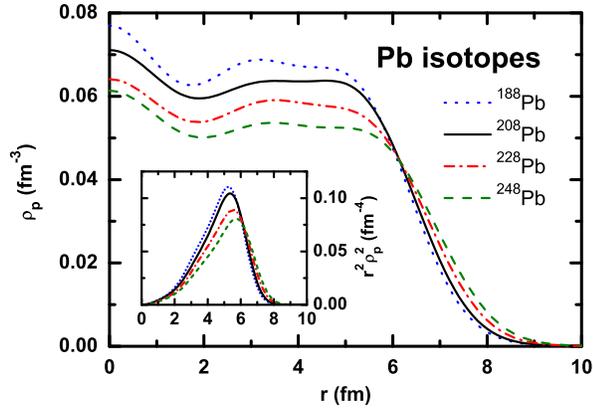}
\caption{(Color online) Proton density distributions $\rho_p(r)$ in $^{188}$Pb, $^{208}$Pb, $^{228}$Pb, and $^{248}$Pb by RHFB theory with PKA1.
The corresponding $r^2 \rho_p^2(r)$ distributions are shown in the insert.
    \label{Fig2}}
\end{figure}

In order to exclude the effects due to self-consistency,
one-step calculations have been performed to investigate the effects
of the LDA approximation on Coulomb exchange energies, i.e., $E^{\rm
RLDA}_{\rm Cex}$ and $E^{\rm NRLDA}_{\rm Cex}$ are respectively
obtained with the relativistic and NR LDA by the same converged
proton density distributions $\rho_p(r)$ given by the
self-consistent RHFB calculations.
By taking the even-even Pb isotopes from proton drip line to neutron
drip line as examples, the Coulomb exchange energies $E_{\rm Cex}$
are shown as a function of mass number $A$ in the upper panel of
Fig.~\ref{Fig1}. The corresponding relative deviations $\Delta
E_{\rm Cex}$ defined in Eq.~(\ref{Eq:DE}) are shown in the lower
panel.
The results calculated by using Eq.~(\ref{Eq:EexNR}), the exact formula for the NR case, are also shown for comparison.
In addition, the proton density distributions $\rho_p(r)$ in
the nuclei $^{188}$Pb, $^{208}$Pb, $^{228}$Pb, and $^{248}$Pb are
illustrated in Fig.~\ref{Fig2}.

In general, the magnitudes of the Coulomb exchange energies $E_{\rm
Cex}$ decrease with increasing mass number, i.e., with increasing
size.
For each nucleus, the magnitude of $E_{\rm
Cex}$ is overestimated by the NR LDA or the traditional Slater
approximation, and substantially improved when the relativistic
corrections are taken into account. From the relative deviations
(see the lower panel of Fig.~\ref{Fig1}), the relativistic
correction improvements range from $4.5\%$ in $^{182}$Pb to $3.5\%$
in $^{266}$Pb.
The isotopic dependence of the relativistic effect in Eq.~(\ref{Eq:ERLDA}) is related to the proton density distributions.
As shown in the insert of Fig.~\ref{Fig2}, it is clear that the relevant $r^2\rho_p^2(r)$ distribution much decreases and becomes slightly narrower close to the surface region with the neutron number, which results in a smaller relativistic correction.
Accordingly,
the relative deviations $\Delta E_{\rm Cex}$ due to the relativistic
LDA decrease from $3.6\%$ in $^{182}$Pb to $1.7\%$ in $^{266}$Pb.
As shown in Fig.~\ref{Fig2}, in the heavy nuclei, the inner proton density is almost flat that forms a near uniform density region, and the tail part of density varies slowly. These are the reasons why the LDA works well.

\begin{figure}
\centering
\includegraphics[width=8cm]{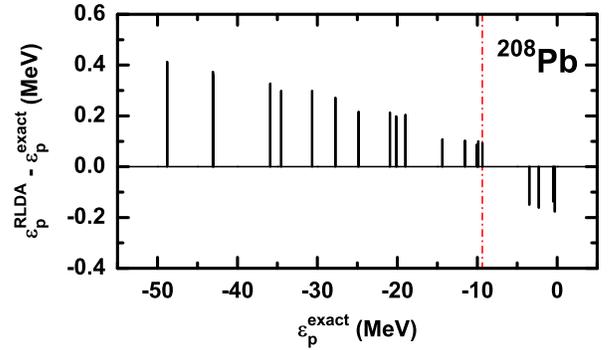}
\caption{(Color online) Proton single-particle energy shifts caused by the RLDA in $^{208}$Pb. The Fermi energy is shown as the vertical dash-dotted line.
    \label{Fig3}}
\end{figure}

In order to investigate the effects due to the self-consistency, the
self-consistent RLDA calculations have been performed. In
Fig.~\ref{Fig3}, the proton single-particle energy shifts
$\varepsilon^{\rm RLDA}_p-\varepsilon^{\rm exact}_p$ in $^{208}$Pb
are shown as a function of the single-particle energy
$\varepsilon^{\rm exact}_p$. It is seen that, in the relativistic
LDA, the occupied and unoccupied proton states are pushed up and
down, respectively. The energy shifts are around $+400$~keV for the
deeply bound states, $+100$~keV for the states near the Fermi
surface, and $-200$~keV for the unoccupied bound states. A very
similar shifting behavior was shown in Ref.~\cite{LeBloas2011}
in the NR Skyrme Hartree-Fock framework. As the occupied states
are pushed up due to the RLDA, the total kinetic energy of protons
increases by $2.73$~MeV, which is completely canceled out by the
self-consistent effects of the neutron kinetic ($-2.12$~MeV), mesons
($-1.18$~MeV), and Coulomb direct ($+0.58$~MeV) energies. The same
kind of self-consistent cancellation between the kinetic energy and
other channels was also noted in Ref.~\cite{LeBloas2011}. As a
result, the total energy difference between the RLDA and RHFB
calculations, $-0.97$~MeV, is mainly due to the difference in the
Coulomb exchange energies, $-0.99$~MeV.

\begin{figure}
\centering
\includegraphics[width=8cm]{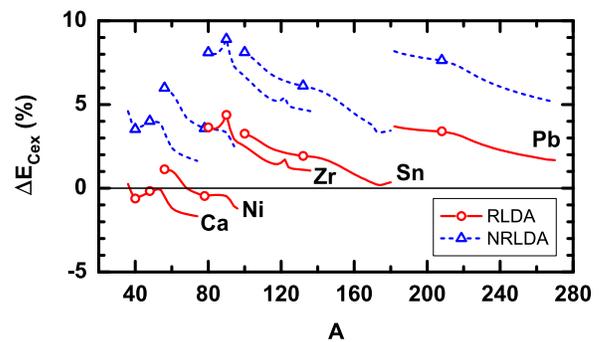}
\caption{(Color online) Relative deviations of the Coulomb exchange energies by the self-consistent RLDA (solid lines) and NRLDA (dashed lines) calculations for Ca, Ni, Zr, Sn, and Pb isotopes. The traditional doubly magic nuclei are denoted as open symbols.
    \label{Fig4}}
\end{figure}

In Fig.~\ref{Fig4}, we show the systematics of calculated results
using the self-consistent relativistic LDA for the semi-magic Ca,
Ni, Zr, Sn, and Pb isotopes from proton drip line to neutron drip
line. The traditional doubly magic nuclei are marked by the open
symbols. Comparing with the exact Coulomb exchange energies
calculated in RHFB, it is found that the relative deviations $\Delta
E_{\rm Cex}$ introduced by the relativistic LDA are less than $5\%$
for all these five semi-magic isotopes. In addition, from the
differences between the solid and dashed lines, the relativistic
corrections to the LDA are found to play substantial roles in
improving the results by $3\sim5\%$.

Therefore, one can conclude that the relativistic version of LDA shown in Eqs.~(\ref{Eq:ERLDA}) and (\ref{Eq:VRLDA}) for the Coulomb exchange term in nuclear CDFT is very robust and promising.
In particular, the relativistic corrections to the traditional Slater approximation shown in Eqs.~(\ref{Eq:exR}) and (\ref{Eq:Phi}) are very important.

Another good test for the RLDA is the Coulomb Displacement Energies (CDE) of mirror nuclei. As one of the clearest cases \cite{VanGiai1971}, it is found that the experimental mirror binding energy difference $E_B(^{17}{\rm O}) - E_B(^{17}{\rm F}) = 3.54$~MeV \cite{Wang2012} can be well reproduced by the RHFB calculation, $3.47$~MeV, with blocking for the odd nucleon \cite{Long2010a}, while $3.33$~MeV by the RLDA calculation. Another representative case is the pair of $^{48}$Ni and $^{48}$Ca \cite{Brown1998}. For this case, the mirror binding energy difference $E_B(^{48}{\rm Ni}) - E_B(^{48}{\rm Ca})$ obtained by the RHFB and RLDA calculations are almost identical, which are $63.44$ and $63.38$~MeV, respectively. Nevertheless, the experimental value $66.95(53)$~MeV \cite{Wang2012} is underestimated to some extents, because the charge-symmetry breaking interactions, which are found important for CDE \cite{Brown1998,Brown2000}, are not included.


In summary, the relativistic LDA for the Coulomb exchange
functional in nuclear systems is presented. This approximation is
composed of the well-known Slater approximation in the NR scheme and
the corrections due to the relativistic effects. The validity of the
relativistic LDA in finite nuclei calculations is examined by
comparing with the results of the RHFB theory, where the non-local
Coulomb exchange term is treated exactly. It is found that the
relative deviations of the Coulomb exchange energies in the
self-consistent RLDA calculations are in general less than $5\%$ for
semi-magic Ca, Ni, Zr, Sn, and Pb isotopes from proton drip line to
neutron drip line. For the proton single-particle energy shifts, the
relativistic LDA pushes the occupied and unoccupied states upward
and downward, respectively, which is in agreement with the previous
NR Hartree-Fock calculations. Finally, it is also worthwhile to
emphasize that the relativistic corrections to the LDA are found to
play substantial roles in improving the agreement with the exact
results by $3\sim5\%$.
This study opens a new door for implementing the Coulomb exchange effects in the RH theory yet keeping its merits of locality in the future.

This work is partly supported by
the Fundamental Research Funds for Central Universities under Contracts No. lzujbky-2011-15 and No. lzujbky-2012-k07,
the Major State 973 Program 2013CB834400,
the National Natural Science Foundation of China under Grants No. 10975008, No. 11075066, No. 11105006, and No. 11175002,
the Research Fund for the Doctoral Program of Higher Education under Grant No. 20110001110087,
the Grant-in-Aid for JSPS Fellows under Grant No. 24-02201,
and the Program for New Century Excellent Talents in University of China under Grant No. NCET-10-0466.


\end{document}